\documentstyle[12pt,preprint]{aastex}
\newcommand{\kms}{km~s$^{-1}$}
\newcommand{\kmsM} {km~s$^{-1}$~Mpc$^{-1}$}
\newcommand{\subsun}{\mbox{$_{\odot}$}}
\newcommand{\etal}{{\it et al.\/}}
\newcommand{\rxj}{RX~J1347$-$1145}

\begin{document}

\title{Losing Weight: A KECK Spectroscopic Survey of the Massive
Cluster of Galaxies \rxj\altaffilmark{1} }

\author  {Judith G. Cohen\altaffilmark{2} \& 
Jean-Paul Kneib\altaffilmark{3} }

\altaffiltext{1}{Based in large part on observations obtained at the
        W.M. Keck Observatory, which is operated jointly by the California 
        Institute of Technology, the University of California
        and NASA,}
\altaffiltext{2}{Palomar Observatory, Mail Stop 105-24,
        California Institute of Technology, Pasadena, CA \, 91125}
\altaffiltext{3}{Observatoire Midi-Pyrenees, 14 Av. E.Belin, 
31400 Toulouse, France}

\begin{abstract}

  We present a sample of 47 spectroscopically confirmed members of 
  \rxj, the most luminous X-ray cluster of galaxies discovered to
  date. With two exceptions, all the galaxies in this sample 
  have red $B-R$ colors and red spectral indices, with 
  spectra similar to old local ellipticals.
  Using all 47 cluster members, we derive a mean redshift of $\bar{z}
  = 0.4509\pm 0.003$, and a velocity dispersion of $910\pm130$ \kms, 
  which corresponds to a virial mass of $4.4 \times 10^{14}
  h^{-1} M$\subsun\ with an harmonic radius of $380 h^{-1}$ kpc.  The
  derived total dynamical mass is marginally consistent with that deduced 
from the cluster's X-ray emission based on the analysis of 
  ROSAT/ASCA images (Schindler \etal\ 1997), but
  not consistent with the more recent X-ray analyses of Allen (2000),
Ettori, Allen \& Fabian (2001) and Allen, Schmidt \& Fabian (2002).
  Furthermore, the dynamical mass is significantly
  smaller than that derived from weak lensing (Fischer \& Tyson
  1997) and from strong lensing (Sahu \etal\ 1998).
  We propose that these various discrepant
  mass estimates may be understood if \rxj\ is
  the product of two clusters caught in the act of merging
  in a direction perpendicular to the line of sight, although there
is no evidence from the galaxy redshift distribution supporting this
hypothesis.  Even with this hypothesis, a significant part of 
the extremely high X-ray
luminosity must still arise from non-virialized, presumably
shocked, gas.
  Finally, we report the serendipitous discovery of a lensed background 
  galaxy at $z=4.083$ which will put strong constraints on the lensing mass determination 
  once its counter-image is securely identified.

\end{abstract}


\section{Introduction}

As the most massive gravitationally bound objects in the Universe,
galaxy clusters are prime targets for studies of structure formation
and evolution. In order to use these massive objects as cosmological
tools, a good understanding of their mass distribution is required
to relate the numerical simulation predictions to the observations.
Because of their generic rarity, the most massive
clusters constitute in principle the most sensitive cosmological probes, 
with the the most distant ones providing the tightest constraints, 
specifically on the value of $\Omega_0$ 
(e.g., Bahcall \& Fan 1998; Ebeling et al.\ 2001). X-ray selection
is currently the most favored technique for finding these massive systems
in the Universe because of the very well defined selection criteria.
However the most extreme clusters may also be the most unusual
cases that may not be present in the current
generation of numerical simulations of 
the Universe. Hence, precise understanding of the most massive clusters 
is very important in order to derive any useful cosmological constraint.

As an example, MS1054--0321, the highest
redshift cluster in the Einstein Medium Sensitivity Survey, 
is a very massive cluster.  Donahue
\etal\ (1998) ascribed a virial mass of $7.4 \times 10^{14} h^{-1}$
M\subsun\ (for $\Omega_m = 1$) to
MS1054--0321 based on both its X-ray temperature and a velocity
dispersion from a sample of twelve spectroscopic members of
$1360\pm450$ \kms.  However, even in this initial study of
MS1054--0321, the presence of considerable substructure was noted.
More recently, a careful weak lensing analysis using HST images by
Hoekstra, Franx \& Kuijken (200), a velocity dispersion analysis with
a larger sample of spectroscopically confirmed cluster members (78
cluster members, giving $\sigma_v=1150\pm97$ \kms) by van Dokkum
\etal\ (2000), and new $Chandra$ observations (Jeltema \etal\ 2001) all
seem to support a value for the virial mass of MS1054--0321 about a
factor of two lower.

We focus in this paper on the cluster of galaxies
\rxj\ ($z$=0.451), discovered by ROSAT. 
This cluster is the most luminous X-ray 
cluster discovered to date
(Schindler \etal\ 1995) with an intrinsic bolometric X-ray luminosity 
\footnote{All published values have been adjusted to the common value
of $H_0 = 100$ \kmsM, and the explicit dependence on $h = H_0/100$
is given.} of $L_{bol} = 50 \times 10^{44} h^{-2}$ ergs s$^{-1}$
and a gas temperature of $T_X = 9.3\pm1.0$ keV from ASCA
observations (Schindler \etal\ 1997).  The presumption that this is a high mass
object is enhanced by the discovery of two reasonably bright long arcs
with lengths of $\sim$6 arcsec (see Sahu \etal\ 1998 for the lensing analysis)
and by a clear detection of weak
lensing (Fischer \& Tyson 1997).  More recently,
Komatsu \etal\ (1999) and Pointecouteau \etal\ (2001) have
respectively tried to estimate the cluster mass by the measure of the
Sunyaev-Zeldovich (SZ) increment (resp. decrement) at submm (resp. mm)
wavelengths. 

We present here Keck spectroscopy of galaxies in \rxj\ 
which allow us to probe the dynamics of this massive cluster.
Section 2 describes the observations and the spectral characteristics
of the 47 cluster members.  We also report the discovery of a 
lensed galaxy at $z=4.083$.  A detailed
analysis of the spectra of the cluster members is carried out using
spectroscopic indices in \S 3.
In \S\ref{section_vdisp}, we discuss the results
in terms of the dynamical estimate of the mass of this cluster, and compare the
derived mass to other mass estimates in \S\ref{section_compare}.  
We derive a possible
model for the mass and dynamics of \rxj\ in \S\ref{section_merger}.
A brief summary and discussion of the
prospects for improving our understanding of
the mass distribution of this cluster are included in the last
section.

For consistency with earlier work, we adopt the cosmology $\Omega_m =
0.3$, $\Lambda = 0.0$ and $H_0=100 h$~km/s/Mpc so that the angular
scale at the distance of \rxj\ is 3.67 $h^{-1}$ kpc/arcsec.  Were we
to adopt a flat Universe with $\Lambda=0.7$, the angular scale would
increase by $\sim 10$\% to 4.04 kpc/arcsec.

\section{Keck Spectroscopy \label{section_spectra}}

Candidate member galaxies in the cluster \rxj\ were
first selected in 1998, as extended objects of appropriate brightness from
a stacked $R$-band image (4$\times$200 sec)
taken with Low Resolution Imaging Spectrograph (LRIS) 
(Oke \etal\ 1995). Objects are identified using their J2000 coordinates,
so that G47345\_4632 has coordinates RA = 13 47 34.5  
and Dec = $-$11 46 32.
Absolute astrometric calibration of the field was carried out using the
USNO-A V2.0 (Monet \etal\ 1998). 

Beginning in early 2000, we used a
color image (made of a $B$ and $R$-band image) to select candidates 
with suitably red $B-R$ color for spectroscopic observations.
This significantly enhanced the efficiency of detecting
cluster members rather than foreground or background galaxies.  However,
it has probably introduced a bias favoring the inclusion
of elliptical galaxies in our sample.
Three slitmasks were used with LRIS, one in 1998 ($t_{exp}=2500$s), one in 2000 ($t_{exp}=2500$s), and one in
May 2001 ($t_{exp}=2\times 2500$s).  The 300 g/mm grating yielding a dispersion of 2.5\,\AA/pixel
with a 1.0 arcsec wide slit (projected width of 4.7 pixels) was used
for all masks.  
The spectra were reduced in a standard way using Figaro (Shortridge
1993) scripts.

Redshifts were determined by centroiding the CaII H and
the K lines [JC], as well as using a cross correlation
technique ({\sl IRAF/RVSAO2.0} package) [JPK].  The results were
indistinguishable within the errors.  From this analysis,
a sample of secure 46 members was isolated, with one additional 
probable cluster member (C47314\_4511)
whose spectrum was too noisy to yield an accurate
redshift.  However, the presence of a detectable 4000\,\AA\ break
strongly suggests that it belongs to the cluster.  Five of the cluster
galaxies have two independent spectra taken in different runs;
these show no systematic redshift offsets.

Table~\ref{table_cluster_mem} gives the location 
and redshifts for the cluster members, with total $R$ magnitudes
from the best fit large aperture results from the FOCAS package
described in Valdes (1989).  Table~\ref{table_cluster_nomem} lists the
non-members observed, consisting of 22 field galaxies and their galaxy
spectral classification according to the system of Cohen \etal\ 
(1999), as well as stars, some of which were used to align the
slitmasks.  Figure~\ref{figure_rimage}
shows the $R$-band image of the central part of \rxj , where we have
indicated  the galaxies in our sample with their redshifts.

The two central cDs have approximately equal $R$-band luminosity,
but display very different spectra.  The Western cD galaxy,
located at the X-ray peak, is
an AGN and has been detected as a radio point source by the NVSS
(Condon \etal\ 1998, Bauer \etal\ 2000).  
Its spectrum shows extremely strong [OII]\,3727\,\AA\ emission 
as well as emission at H$\beta$. Weaker emission at [OII] at 4959 and
5007\,\AA\ as well as H$\gamma$ and perhaps H$\delta$ is detected.
These emission lines have been observed in the spectra of many giant elliptical
galaxies sitting  at the center of cooling flow clusters ({\it e.g.}
Crawford \etal\ 1995).

The galaxy C47229\_4519
has a spectrum very different from any other
cluster member.  It is, excluding the Western cD, the only cluster galaxy 
showing even weak [OII] 3727\,\AA\ emission, and
its redshift ($z=0.4392$) is slightly lower than all
but one other cluster member.  It is located in the outer part of
\rxj\ and has a blue $B-R$ color.  We include it as a cluster member.

Four galaxies (C47229\_4519 included) 
are slightly offset
in redshift compared to the main cluster redshift distribution.
Perhaps they
are either linked to the cluster infall region or are part of the
large scale structure surrounding the cluster.

\begin{figure}
\epsscale{0.8}
\plotone{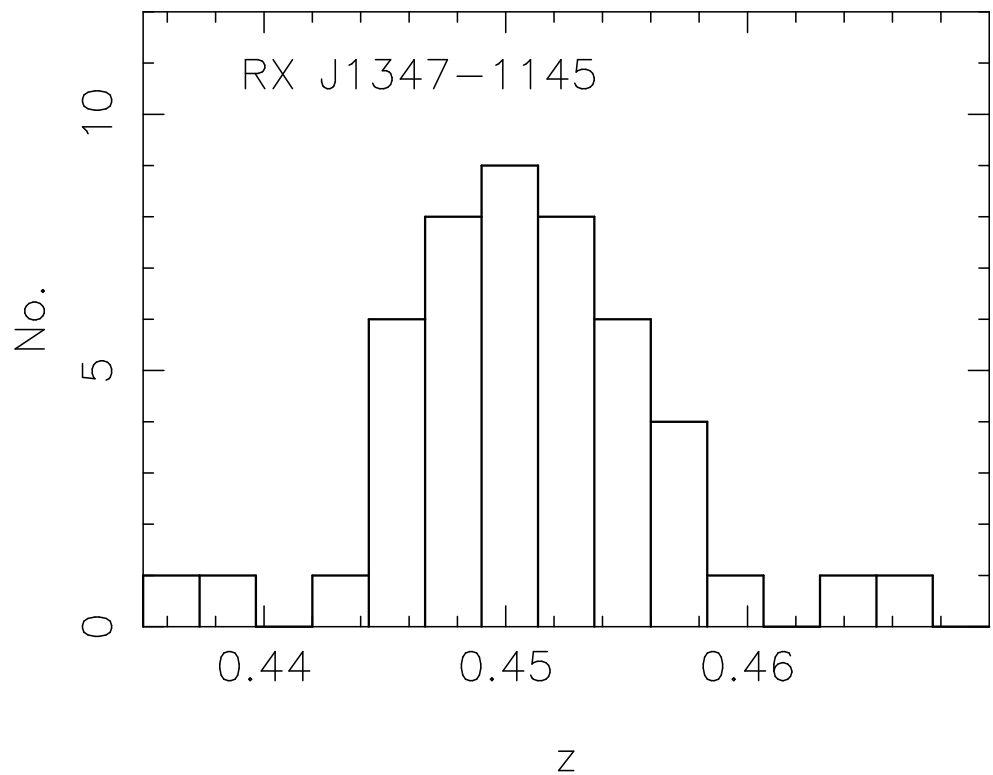}
\caption[f2.eps]{The histogram of velocities for 
47 spectroscopically confirmed members of  the massive cluster
of galaxies \rxj\ is shown.    There
are no other galaxies in the sample with $0.41<z<0.52$.
\label{figure_vel_disp}}
\end{figure}

Altogether, the total number of spectroscopically confirmed galaxies
in the close vicinity of \rxj\ is 47.  Figure~\ref{figure_vel_disp}
shows the velocity distribution centered on the cluster redshift. Note that
there are no other galaxies in the spectroscopically observed sample
with $0.41<z<0.52$.

\subsection {Detection of a Lensed Object at $z=4.083$}

\begin{figure}
\epsscale{1.0}
\plotone{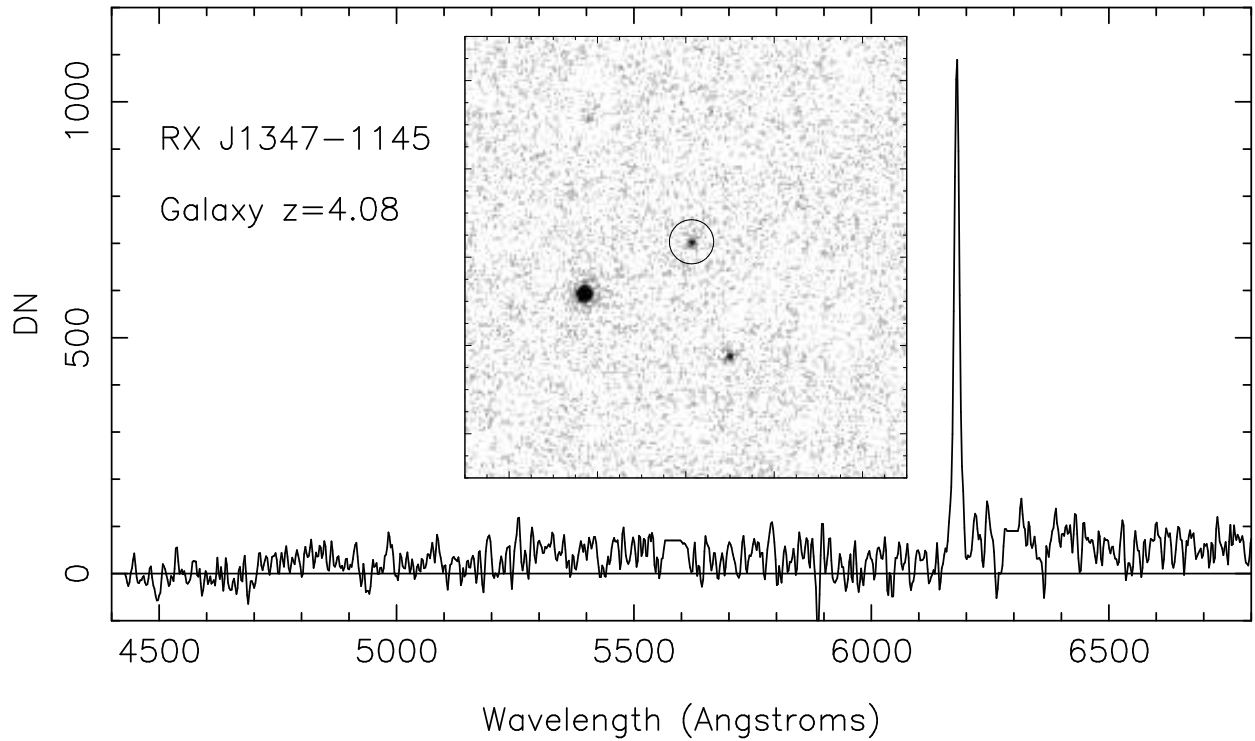}
\caption[f3.eps]{The spectrum of the $z=4.083$ galaxy found near the
center of the cluster of galaxies \rxj\ is shown.  The residuals from
subtraction of the strong night sky lines at 5577\,\AA\ and at 6300\,\AA\
have been removed by setting the counts to a constant within those 
specific intervals.  The superposed image is a zoom on the archival 
HST/STIS data showing
the point source morphology of the object.
\label{figure_z4spec}}
\end{figure}

During the spectroscopic survey to locate members of the cluster of
galaxies \rxj, we serendipitously found an object (O47332\_4511)
with a strong emission line at 6177\,\AA\ and essentially no 
continuum blue-ward of this line. Identifying the line with 
Lyman-$\alpha$ and the lack of blue
continuum as the Lyman limit
gives a redshift of $z=4.083$ (Figure~\ref{figure_z4spec}),
and the equivalent width of Lyman-$\alpha$ is
$\sim$45 \AA.  This
object, with $R=23.7$ (although the very strong emission falls within the
$R$ bandpass, it should not contribute to the $R$-band flux by more than 10\%) 
is 38 arcsec East of the Western cD. Its image
is a point source on the STIS image from the
HST Archive (Figure~\ref{figure_z4spec}).

\section{Spectral Indices}

\begin{figure}
\epsscale{0.8}
\plotone{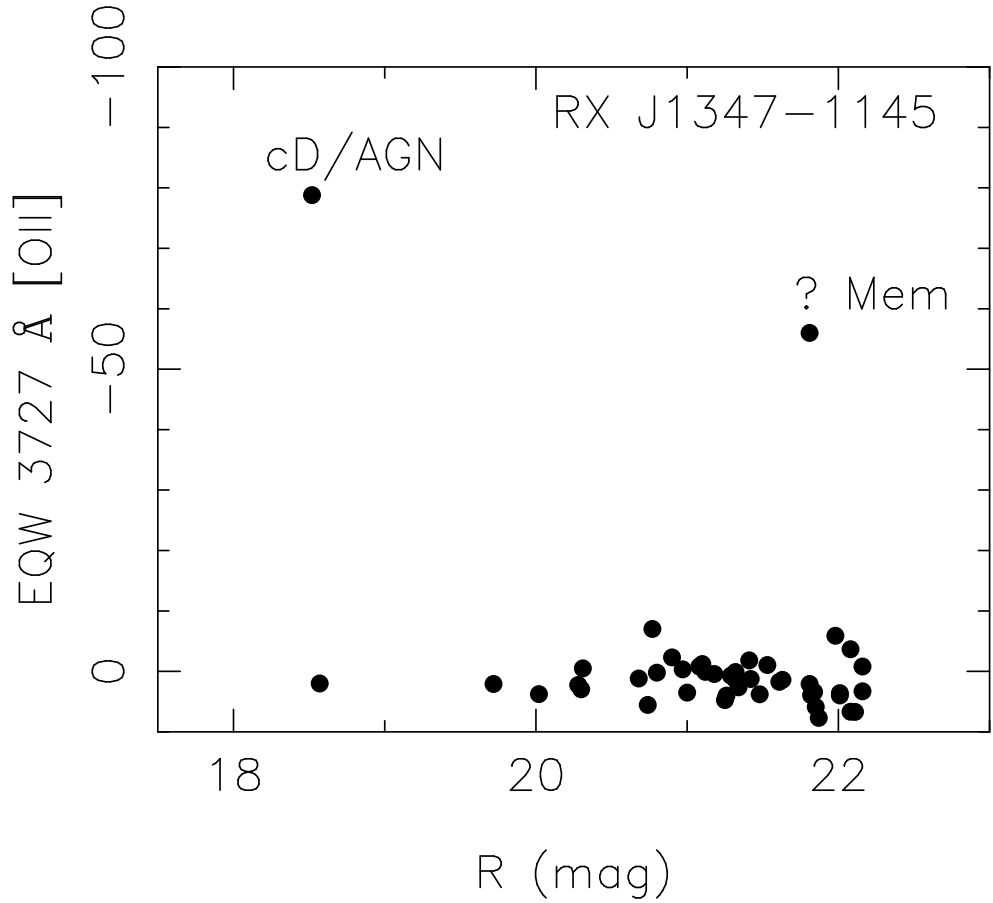}
\caption[f4.eps]{The rest frame equivalent width of the emission
line of [OII] at 3727\,\AA\ is shown as a function of total $R$ mag
for the sample of galaxies in the massive cluster \rxj.
The central AGN and one possible non-member, marked on the figure,
are the only galaxies in the
sample with detectable emission in this line.
\label{figure_3727}}
\end{figure}

\begin{figure}
\epsscale{0.8}
\plotone{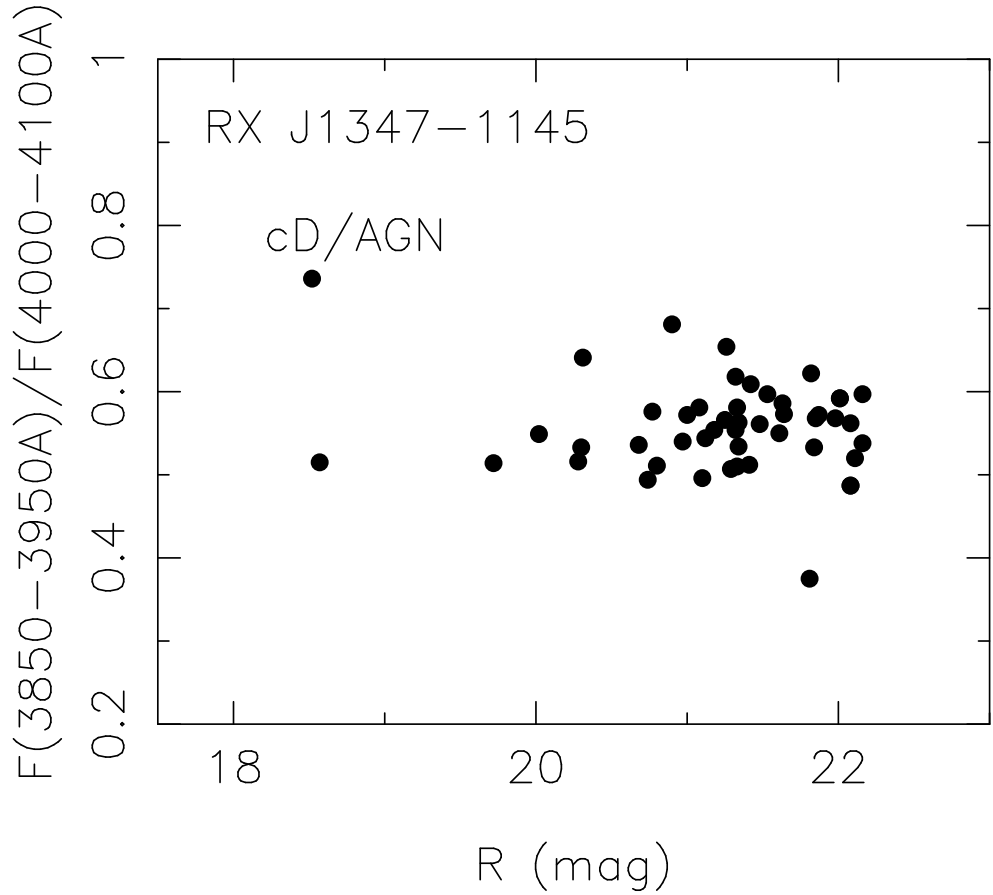}
\caption[f5.eps]{An index measuring the strength of the
Balmer jump in the rest frame is shown as a function of total $R$ mag
for the sample of galaxies in the massive cluster \rxj.
The value 1.0 corresponds to no discontinuity in the spectrum.
The central AGN and one possible non-member, which has such a blue
continuum that its Balmer discontinuity index exceeds unity,
are the only anomalous galaxies in this plot.
\label{figure_baljump}}
\end{figure}

\begin{figure}
\epsscale{0.8}
\plotone{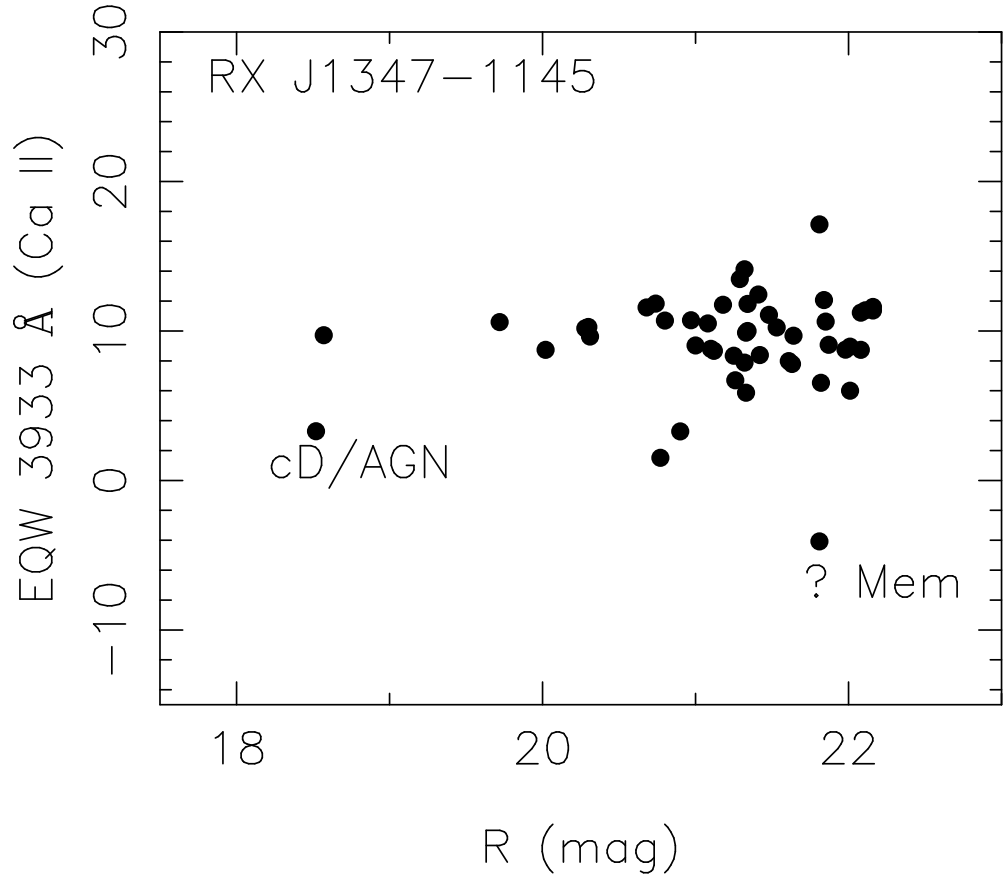}
\caption[f6.eps]{The rest frame equivalent width of the Ca II
absorption line at 3933\,\AA\
is shown as a function of total $R$ mag
for the sample of galaxies in the massive cluster \rxj.
The central AGN and one possible non-member, marked on the figure,
have anomalously weak absorption in this line.
\label{figure_3933}}
\end{figure}

To illustrate that the members of the cluster of galaxies \rxj\ are
typical early type galaxies, we have measured the strengths of various
spectral features, the strength of emission in the 3727\,\AA\ line of
[OII], and the absorption in the 3968\,\AA\ line of CaII as well as an
index for the strength of the 4000\,\AA\ break.  Table~\ref{table_bands}
lists the specific bandpasses used to define each of these indices.
The results as a function of total $R$ mag are shown in
Figures~\ref{figure_3727}, \ref{figure_baljump} and \ref{figure_3933}.
The possible non-member galaxy C47229\_4519
is included in each of
these figures.

Aside from the AGN and the possible non-member galaxy, the remaining
45 members of \rxj\ show no evidence for emission in the 3727\,\AA\ line
of [OII] (see Figure~\ref{figure_3727}).  Repeat measurements 
suggest an accuracy of $\pm3$\AA\ for
the fainter galaxies in our sample.  The same two galaxies stand out
from the rest of the sample in Figure~\ref{figure_baljump}, where they
display Balmer jumps considerably smaller than the majority of the
cluster galaxies, and in Figure~\ref{figure_3933}, where they have
less absorption in the 3933\,\AA\ line of Ca II.  All of this is to be
expected given that one of these galaxies in an AGN and the second
appears to be very blue, presumably from ongoing star formation.

\section{Velocity Dispersion and Virial Mass \label{section_vdisp}}

The velocity dispersion was computed using the bi-weight algorithm of
Beers, Flynn \& Gebhardt (1990), as it is very robust to the presence
of outliers.  An instrumental uncertainty of $\pm100$ \kms\ in the
rest frame is assumed for all measurements.  The bi-weight estimator
gives $\sigma_v = 910\pm130$ \kms\ using the 47 known members of the
cluster \rxj.  A slightly smaller value ($\sigma_v = 820\pm110$ \kms) 
is obtained using a classical
3-sigma clipping algorithm (which has the effect of removing the four
outlier galaxies discussed in \S\ref{section_spectra}, which are
slightly offset from the main distribution).  Hereafter, we adopt a velocity dispersion of 910 \kms\ for this cluster.  
Assuming that the cluster follows the $\sigma$--$T_X$ relation ({\it e.g.}
Girardi \etal\ 1996), we predict an X-ray temperature of
$T_X= 5.1\pm1.2$ keV.

The central redshift for the velocity distribution is $z=0.45095$.
The redshift of the Eastern and Western cDs are respectively
$z=0.4506$ and $z=0.4515$, both of which correspond to rest frame radial
velocities that are within 100 \kms\ of the central value.  Therefore,
taking into account the measured uncertainties, they are both
consistent with being at rest at the dynamical center (in the line of sight
direction); however we can not speculate on any transverse
velocities.

The centroid of the galaxy distribution appears to be closer to the
Western cD, rather than to the point half way between the two cDs,
which are 18 arcsec apart. Thus, we adopt the position of the AGN as the
dynamical center of the cluster, as it is also corresponds to the X-ray
peak.

As Figure~\ref{figure_vel_disp} shows, the velocity distribution
appears to be that of a Gaussian.  A K-S test shows that the probability
that the observed velocity distribution for the members of
the cluster \rxj\ and the fit Gaussian are the same exceeds 98\%.
There is no evidence  in our sample of any deviation of the distribution
of the radial velocities from a single Gaussian.

We computed the harmonic radius $R_h$ (e.g., Saslow 1985,
Nolthenius \& White 1987)
for our spectroscopic cluster members as
\begin{equation}
\displaystyle
R_h = D_A(\bar{z}) {\pi\over 2}{N_m(N_m-1)\over 2}
\left(\Sigma_i\Sigma_{j>i}\theta_{ij}^{-1}\right)^{-1},
\end{equation}
where $\theta_{ij}$ is the angular distance between galaxies $i$ and
$j$, $N_m$ is the number of cluster members, and $ D_A(\bar{z})$ is
the angular diameter distance at the mean cluster redshift $\bar{z}$.
The cluster virial mass can then be estimated as
\begin{equation}
M_V = { 6 \sigma^2 R_h \over G}.
\end{equation}
We found an harmonic radius of $R_h = 380 h^{-1}$ kpc and a mass
$M_V = 4.4^{+1.4}_{-1.2}{\times}10^{14} h^{-1}$ M$_\odot$.  
This value is somewhat
larger than that derived from the measured $\sigma(v)$ alone using the
fits of Arnaud \& Evrard (1999) to the simulations of Evrard \etal\ 
(1996) for the mass within a region whose density is 200 times the
critical density, $M = 2.9 \times 10^{14} h^{-1} M$\subsun.

The fact that
most members of our sample of galaxies in \rxj\
are red ellipticals argues that the dynamical estimate is
a secure estimate of the cluster's mass.

\section{Comparison of the Virial Mass With Other Mass Estimators
\label{section_compare}}

Measurements of cluster masses deduced from X-ray emission are based
on the assumption that the X-ray emitting gas is in hydrostatic
equilibrium with the gravitational potential of the cluster. The 
constant $T_X=9.3\pm 1.0$ keV temperature measured from ASCA data
by  Schindler \etal\ (1997) corresponds, using
the $T_X-\sigma$ relation of Girardi \etal\ (1996) (as we do
throughout), to a velocity dispersion of $1320\pm100$ \kms.  With
this $T_X$,
Schindler \etal\ (1997) deduced from the
ROSAT images of \rxj\ a gas mass of $1.0 \times 10^{14}$ h$^{-1}
M$\subsun\ and a total binding mass of $2.9 \times 10^{14}$ h$^{-1}
M$\subsun\ within a radius of 500 $h^{-1}$ kpc.

X-ray mass determinations are usually given as a mass enclosed
within a specified radius.  We adopt a power law 
$\rho(r) \propto r^{-x}$ for the spatial
distribution of galaxies within a cluster to determine the conversion between
the harmonic radius and the outer radius, $r_{max}$. 
Kent \& Gunn (1982) found $x \sim 3$ is the appropriate power law
to characterize the distribution of galaxies in the 
Coma Cluster, and we adopt that value.  
Ignoring a central core extending to $r = 0.05r_{max}$,
the harmonic radius is then
0.49$r_{max}$, and this fraction decreases for an even 
steeper density drop-off or as more of the core is included.  
Thus the area surveyed by our optical sample, with $R_h = 380 h^{-1}$ kpc,
is roughly comparable to an X-ray mass specified within 1$h^{-1}$ Mpc.

\rxj\ has a large cooling flow ($\gtrsim1000$ M\subsun/year).  Its
high luminosity and mass are confirmed by the recent analysis of
BeppoSAX observations of this cluster by Ettori \etal\ (2001), who
focus on the cluster gas temperature and the impact of the large
cooling flow in its central region.  They find, confirming Allen
(2000), a gas temperature of ${\sim}14$ keV, but not as large as the
very high value of 26.4$^{+7.8}_{-12.3}$ keV of Allen (1998).

The recent $Chandra$/ACIS \rxj\ image
(a 20 ksec exposure we extracted from the $Chandra$ archive) shows an
extended source whose center coincides with the AGN (the Western of
the two central cDs) to within 0.8 arcsec (the relative astrometry
was checked using the only X-ray point source overlapping with our
LRIS image, which
was identified with a faint field galaxy). Taking into account all
the astrometric uncertainties, the Western cD and the X-ray peak are consistent
with both being at exactly the same position.
The contribution of the AGN
itself to the total X-ray flux is likely to be small.  
A more detailed analysis of these data by Allen, Schmidt \& Fabian (2002)
leads again to a very high X-ray temperature, $T_X = 12.0\pm0.6$ keV,
which corresponds to $\sigma(v) = 1545\pm50$ \kms.

The SZ results of Pointecouteau \etal\ (2001)
lead to projected gas mass of $1.9\pm0.1 \times 10^{13}$ h$^{-5/2}$
M\subsun\ within an angular radius of 74 arcsec (272 $h^{-1}$ kpc)
assuming a spherical
distribution for the gas.  This value is in reasonable agreement with
the X-ray determination for the gas mass in this cluster.

The virial mass of \rxj\  
can be compared to the mass derived from
weak lensing of Fischer \& Tyson (1997) of $1.1\pm0.3 \times 10^{15}
h^{-1} M$\subsun\ within the same 1 $h^{-1}$ Mpc radius for this
cluster.  
Assuming an isotropic
velocity distribution, the weak lensing results translate into a
predicted velocity dispersion of $1500\pm160$ \kms\ (Fischer \& Tyson
1997).  

The strong lensing analysis for one of the central arcs by
Sahu \etal\ (1998) also suggests a very high mass.\footnote{We have 
confirmed their redshift for the brightest arc, but failed to
obtain any credible redshifts for the fainter arcs in \rxj.}   
They derive the projected mass within the radius of the arcs, 
38 arcsec (140 $h^{-1}$ kpc), and
obtain $3.4 \times 10^{14} h^{-1} M$\subsun, which corresponds to
roughly $\sigma=$1300 \kms\ for a singular isothermal sphere. 
This mass estimate assumes that the
$z=0.81$ arc seen in \rxj\ is located at the Einstein radius, but 
no multiple images were identified
in this analysis. Thus the derived strong lensing mass estimate should
be considered as an upper limit.

We thus see that our measured velocity dispersion
for the X-ray luminious and massive cluster of galaxies \rxj\ 
is significantly smaller than that
inferred from the X-ray or SZ measurements or from
both weak and strong lensing studies.
Table~\ref{table_sigmas} summarizes the current discrepant situation,
in terms of the velocity dispersion, X-ray temperature or total projected
mass within some specified radius.
The various measurements have been converted assuming
the $T_X-\sigma$ relation of Girardi \etal\ (1996); the values
so derived are given in brackets in columns 3 and 4 of this table.

\subsection {The Lensed Object at $z=4.083$}

\begin{figure}
\epsscale{1.0}
\plotone{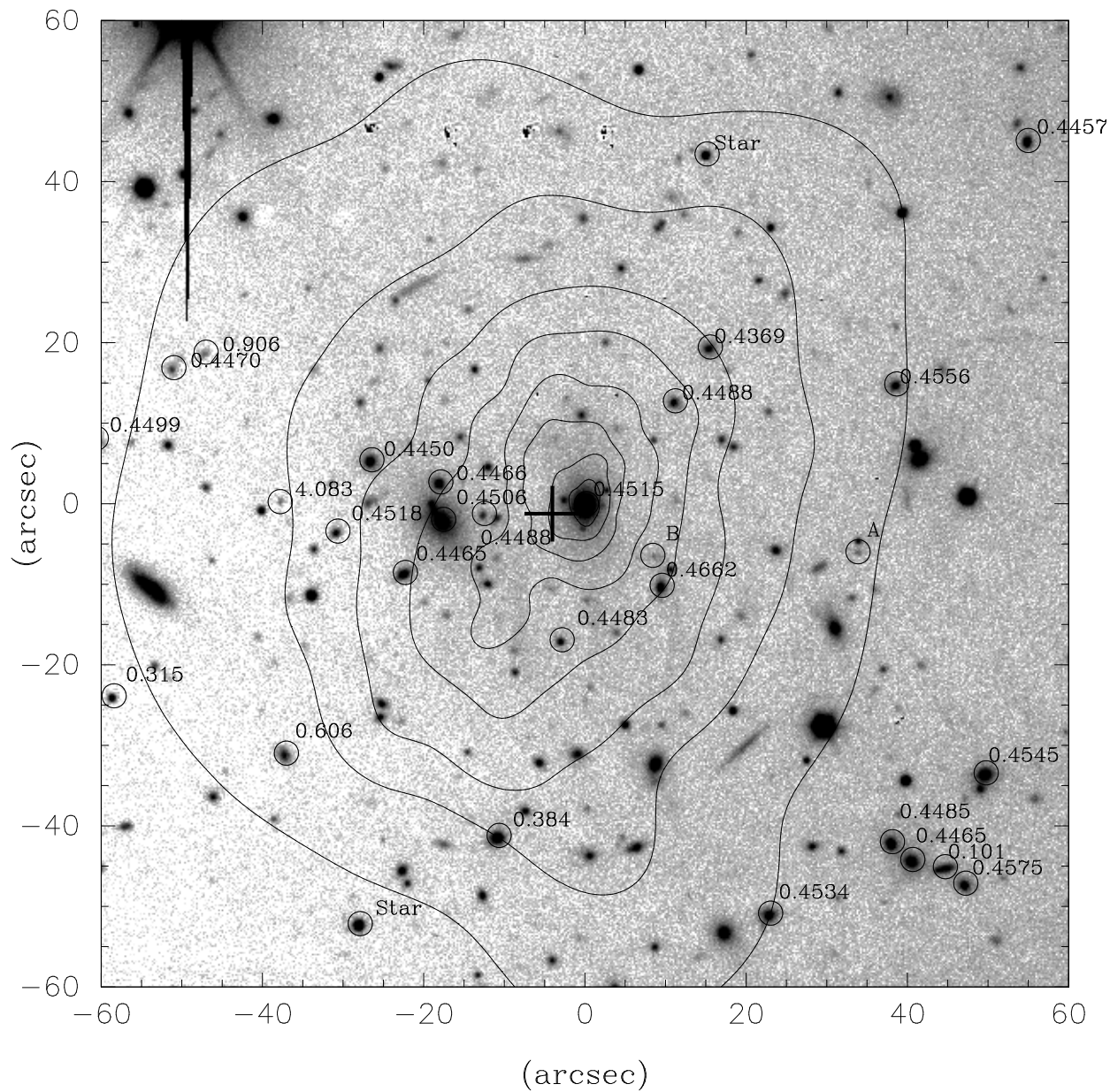}
\caption[f1.eps]{The $R$-band image of the central $2\times 2$ arcmin$^2$
region of \rxj\ is shown, with the $Chandra$ smoothed X-ray surface-brightness
contours (logarithmically spaced) overlaid.
The objects in our spectroscopic sample are marked with their
redshifts. We also marked the possible counter images (A and B) 
of the $z=4.083$
object.
\label{figure_rimage}}
\end{figure}

We attempt to use the lensed high redshift galaxy we have discovered
in this cluster to constrain its mass.
With the aid of simple lensing mass models (following the precepts
of Kneib \etal\ 1996), we have
identified two faint objects with the appropriate 
morphology that might be counterparts of this $z=4.083$ object:
O47283\_4517 
and O47300\_4517 (objects A and B respectively, marked in Fig.~\ref{figure_rimage}).
These models assume that the $z=4.083$ object is multiply imaged 
by the cluster, but this needs to be confirmed in the future; at present
we lack the necessary color and spectroscopic information for
objects A and B.
If we accept either of these two objects marked in Figure~\ref{figure_rimage} 
as a possible counter-image of the $z=4.083$ object, a strong
lensing analysis suggests a much smaller total projected
mass within a 38 arcsec (140 $h^{-1}$ kpc) radius 
centered on the Western cD compared to the previous lensing analysis.  We 
find a mass of $1.4 \times 10^{14}$ M$_\odot$ for O47300\_4517 (which would
correspond to $\sigma\sim$ 850 \kms\ for a singular isothermal sphere)
and $1.9 \times 10^{14}$ M$_\odot$ for O47283\_4517 
($\sigma\sim$ 1000 \kms\ for a similar model).
Both mass models are made of two massive cluster scale components centered
respectively on each of the two central cDs.
The latter case (object A) is more likely to be correct, as in the former, the
observed flux ratio between the two images is different than the one predicted
by the lens model. Both of these mass estimates are much smaller than
the previous weak or strong lensing estimates and both are consistent
with our dynamical estimate.

\section{\rxj\ Viewed As A Merger \label{section_merger}}

We are quite confident of our measured velocity dispersion for the galaxies
in \rxj.  However, as discussed in \S\ref{section_compare},
it is considerably smaller than would be expected
from the X-ray, SZ, or lensing analyses of this cluster of galaxies.
If we accept all published measurements of $T_X$, SZ decrements
and lensing shears as valid, the only way we see to reconcile all the
data is for \rxj\ to be involved in a major merger in the plane of the
sky (hence barely affecting the dynamical mass estimate).
The probability that such a collision might occur is 
proportional to the solid angle
subtended by a collision ``primarily in the plane of the sky'',
which, for relative velocity vectors
$\pm30^\circ$ from the plane of the sky, is
50\%.

A merger with this geometry could 
explain the origin of the various discrepant temperature measurements for this
cluster.
For multiple merging clumps along a line of sight, 
the weak lensing signal adds linearly.  Hence, irrespective of
whether \rxj\ is in the process of a merger, 
the weak lensing mass estimate should yield the correct total mass
for the cluster.  
The dynamical mass estimate, however,
will be biased toward the mass of the larger clump, at least
until the merger is complete and the galaxy orbits have virialized to
the new total cluster mass.  Given the substantial uncertainties on
the weak lensing mass measurement of Fischer \& Tyson (1997), 
it is just possible to
reconcile the values given in Table~\ref{table_sigmas}
for the dynamical and weak lensing
mass if the masses of the two hypothetical clumps 
($M_1$ and $M_2$, with $M_2 \le M_1$) are comparable, with
$0.7 < M_2/M_1 \le 1$.

Turning to the X-ray measurements,
the observed
values of $T_X$ are very high compared to our dynamical 
galaxy velocity dispersion.  
It is likely that the X-ray emitting gas virializes more rapidly in the course
of a merger than do the galaxies, and hence might reflect the
new total mass of the cluster  before
the galaxy velocity dispersion would do so.    To reproduce 
the observations with a merger hypothesis, we require
equal mass 
clumps\footnote{We assume that the probability of a merger involving
three or more equal mass clumps is so low that we can ignore such events.},
a merger primarily in the plane of the sky,
and also a time chosen so that the X-ray gas has virialized to the
new cluster total mass but the 
galaxies have not.  Even with this somewhat contrived scenario,
it is still not possible to reconcile
our dynamical mass with the most recent X-ray analysis (Allen \etal\ 2002)
unless the optical velocity dispersion has been underesimated by
at least  2$\sigma$, while
$T_X$ has been overestimated by 2$\sigma$. Many of the other recent X-ray analyses
of \rxj\ yield even higher values of $T_X$.

Continuing under the assumption that all published measurements
are correct, we see that
substantial X-ray emission from gas that has
not yet virialized, presumably
from shocks associated with the hypothesized merger, 
is still required to reproduce the  X-ray emission of \rxj.
Theoretical support for such is given by
the hydrodynamic simulations of Ritchie \& Thomas (2001), who
demonstrate that X-ray luminosity and/or temperature
may be strongly enhanced in merging clusters.  

Another clue often used to detect mergers in clusters of
galaxies is the presence of substructure.  Among local clusters, for
example, Schuecker \etal\ (2001)
find that $\sim$50\% of a sample of the nearest clusters of galaxies
show evidence for substructure, presumably arising from recent
mergers, based on their ROSAT images.  Turning to dynamical studies,
the Coma cluster (Colless \& Dunn 1996) and
the Cl0024+1654 cluster, studied in detail by 
Czoske \etal\ (2001), among many others, both show evidence of substructure.  
However, in the above two cases, structure is detected within the
velocity (redshift) distribution which is several times larger than the
upper limit we assign to that of the galaxies in \rxj.
The lack of velocity structure
in the velocity histogram of \rxj\ strongly suggests that there is no
substructure in the line of sight direction.

There is, however, some evidence for substructure in \rxj.
The most recent X-ray map (from ACIS/$Chandra$,
shown as an overlay in Figure~\ref{figure_rimage}) is to first
order circularly symmetric with a probable extension to the SE
and extends to a radius $\lesssim$240 arcsec, a region comparable in size to
that of the optical spectroscopic sample presented here. 
(See Allen \etal\ 2002 for a more detailed discussion.) 
A similar SE extension is seen in the SZ decrement map of 
Pointecouteau \etal\ (2001); it is in fact the most prominent peak of
the SZ map, arguing for a very high temperature for this clump
as well as in the map of Komatsu \etal\ (2001).

\section{Summary \label{section_summary}}

Based on 47 spectroscopically confirmed cluster members, 
we have determined the virial mass of \rxj, the most luminous
X-ray cluster known. This mass estimate is much lower than the
most recent X-ray, SZ and lensing mass estimates. Note
that the case of \rxj\ is not unique and that a number of massive
clusters are poorly understood.
All methods of determining the mass of a cluster of galaxies suffer some bias,
which if understood should allow us to understand cluster physics.

In order to reconcile all the published data on \rxj, we suggest that this
cluster is undergoing a major merger primarily in the plane of the sky.
We further suggest that the extremely high X-ray
luminosity of this cluster does not denote an extremely high mass.

This merger assumption, although attractive,
needs to be confirmed. As indicated above, in our present
sample of 47 spectroscopically confirmed cluster members, there
is no evidence for a merger. 
A detailed 3D analysis of the cluster dynamics
would require a significantly larger sample of spectroscopically
confirmed members of the cluster, quite difficult to obtain.  If
deep HST/ACS images are obtained, we will be able verify
candidate counter-images and
to model accurately the lensing 
distortion  and multiple images to provide additional constraints on
the mass distribution within \rxj.

Large scale programs to search for distant clusters of galaxies are
underway using both X-ray and optical techniques to find distant
clusters.  We need to understand how to measure the mass of clusters
accurately before the comparison of cluster samples at low and high
redshift can be used to constrain cosmology with any degree of
precision.  The massive, X-ray luminous cluster \rxj\ is a perfect
case to try to understand how to do this.

\acknowledgements The entire Keck/LRIS user community owes a huge debt
to Jerry Nelson, Gerry Smith, Bev Oke, and many other people who have
worked to make the Keck Telescope and LRIS a reality.  We are grateful
to the W. M. Keck Foundation
for the vision to fund the construction of the W. M. Keck
Observatory.    The authors wish to extend
special thanks to those of Hawaiian ancestry on whose sacred mountain
we are privileged to be guests.  Without their generous hospitality,
none of the observations presented herein would
have been possible.
The archival STIS data was retrieved from the STScI archive and was taken
with the NASA/ESA Hubble Space
Telescope, which is operated by STScI for the Association of
Universities for Research in Astronomy, Inc., under NASA contract
NAS5-26555.  The archival ACIS/$Chandra$ data was retrieved from the
$Chandra$ Data Archive, which is part of the $Chandra$ X-Ray Science
Center, operated for NASA by the Smithsonian Astrophysical Observatory.
We are grateful to Oliver Czoske, Phil Fisher and Piet van Dokkum 
for useful discussions and suggestions. JPK thanks the CNRS for support.

\clearpage

%
%
\begin{deluxetable}{crr | crr } 
\tablenum{1} 
\tablewidth{0pt} 
\tablecaption{Properties of Members of \rxj} 
\tablehead{
\colhead{Galaxy ID\tablenotemark{a}} & \colhead{$z$}
&
\multicolumn{1}{c|}{$R$} 
&
\colhead{Galaxy ID\tablenotemark{a}} & \colhead{$z$} 
&\colhead{$R$}
\nl 
&\colhead{} 
& 
\multicolumn{1}{c|}
{(Mag)}  
& &
\colhead{}& 
\colhead{(Mag)} 
\nl 
} 
\startdata 
C47223\_4714 & 0.4494 & 21.98 &
C47230\_4432 & 0.4575 & 21.82 \nl 
C47236\_4646 & 0.4560 & 21.34 & 
C7238\_4437 & 0.4503 & 22.16 \nl 
C47243\_4419 & 0.4410 & 22.16 & 
C47251\_4429 & 0.4532 & 21.61 \nl 
C47251\_4556 & 0.4460 & 21.32 & 
C47254\_4530 & 0.4490 & 21.29 \nl 
C47261\_4521 & 0.4502 & 21.25 & 
C47265\_4528 & 0.4604 & 21.10 \nl 
C47268\_4342 & 0.4480 & 21.82 & 
C47269\_4424 & 0.4457\tablenotemark{b} & 21.34 \nl 
C47272\_4543 & 0.4545 & 20.28 &
C47274\_4556 & 0.4575 & 21.08 \nl 
C47278\_4553 & 0.4465 & 20.31 & 
C47280\_4551 & 0.4485 & 20.68 \nl 
C47280\_4454 & 0.4556\tablenotemark{b} & 21.18 & 
C47290\_4600 & 0.4534 & 20.90 \nl 
C47296\_4450 & 0.4369\tablenotemark{b} & 21.32 & 
C47299\_4456 & 0.4488 & 21.63 \nl
C47300\_4519 & 0.4662 & 21.42 & 
C47306\_4509\tablenotemark{c} & 0.4515 & 18.52 \nl 
C47307\_4319 & 0.4526 & 21.26 & 
C47308\_4526 & 0.4483 & 22.11 \nl 
C47314\_4511 & ...\tablenotemark{e} & 21.64 &
C47315\_4510 & 0.4488 & 21.86 \nl 
C47318\_4511\tablenotemark{d} & 0.4506 & 18.57 & 
C47319\_4507 & 0.4466 & 20.74 \nl 
C47319\_4616 & 0.4533 & 22.01 &
C47321\_4352 & 0.4510 & 20.80 \nl 
C47322\_4518 & 0.4465 & 21.33 & 
C47323\_4709 & 0.4550 & 21.53 \nl 
C47324\_4350 & 0.4646 & 20.30 & 
C47324\_4504 & 0.4450 & 20.77 \nl 
C47327\_4513 & 0.4518\tablenotemark{b} & 22.01 & 
C47328\_4614 & 0.4460 & 21.00 \nl 
C47341\_4452 & 0.4470\tablenotemark{b} & 22.08 & 
C47348\_4501 & 0.4499 & 20.97 \nl 
C47357\_4502 & 0.4474 & 21.48 &
C47369\_4434 & 0.4481 & 21.12 \nl 
C47375\_4447 & 0.4512 & 21.33 & 
C47382\_4444 & 0.4495 & 19.72 \nl 
C47384\_4435 & 0.4523 & 21.41 & 
C47395\_4428 & 0.4579 & 21.84 \nl 
C47400\_4533 & 0.4443 & 22.08 & 
C47408\_4523 & 0.4528 & 21.87 \nl 
C47417\_4449 & 0.4557 & 20.02 & 
C47232\_4518\tablenotemark{f} & 0.4392 & 21.81 \nl 
\enddata 
\tablenotetext{a}{Galaxy names are based on their coordinates, 
Cxxyyy\_wwzz has the position 13 xx yy.y $-$11 ww zz, epoch J2000.}
\tablenotetext{b}{There are two independent spectra for this galaxy.} 
\tablenotetext{c}{This is the Western of the two central cDs.} 
\tablenotetext{d}{This is the Eastern of the two central cDs.} 
\tablenotetext{e}{Galaxy on extreme edge of slitlet. Spectrum shows 
this is a cluster member, but redshift not sufficiently accurate 
to use for vel. dispersion.} 
\tablenotetext{f}{Probable member.} 
\label{table_cluster_mem} 
\end{deluxetable}

%
%
\begin{deluxetable}{crrr | crrr } 
\tablenum{2} 
\tablewidth{0pt} 
\tablecaption{Redshift for Non-Members In This Field} 
\tablehead{
\colhead{Galaxy ID\tablenotemark{a}} & \colhead{$R$}
&
\colhead{$z$} &
\multicolumn{1}{c|}{Spec. Type\tablenotemark{b}} 
&
\colhead{Galaxy ID\tablenotemark{e}} & \colhead{$R$} 
&\colhead{$z$} & \colhead{Spec.Type\tablenotemark{b}}
\nl 
& \colhead{(Mag)} 
& \colhead{} &
\multicolumn{1}{c|}
{}  
& &
\colhead{(Mag)}& 
\colhead{} 
\nl 
} 
\startdata 
%
O47234\_4513  & 20.10 & 0.253 & $\cal{E}$,$\cal{I}$ &   
O47240\_4633 &  20.99 & 0.614 & $\cal{EI}$ \nl
O47244\_4604 & $>23$ & 0.607 & $\cal{E}$ &
O47259\_4441 & 20.38 & 0.695 & $\cal{I}$ \nl
O47265\_4528  & 20.92  & 0.299 & $\cal{A}$ & 
O47274\_4351  & 20.88  & star  & $\cal{M}$ \nl
O47276\_4555  & 21.34 & 0.101 &  $\cal{E}$ & 
O47291\_4329 &  21.49 & star & $\cal{M}$ \nl
O47296\_4426 &  21.28 &  star & $\cal{M}$ &
O47314\_4551  &  20.53  & 0.384 & $\cal{A}$ \nl
O47326\_4602  &  20.90  & star &  $\cal{M}$\tablenotemark{c} & 
O47332\_4511 & 23.7 & 4.083 & $\cal{E}$ \nl
O47332\_4540 &  21.52 & 0.606 & $\cal{I}$ &
O47335\_4623  & 20.68 & star & $\cal{S}$ \nl
O47335\_4713  & 21.30   &  star & $\cal{M}$ &
O47339\_4451 & 23.24 & 0.906 & $\cal{E}$ \nl
O47346\_4533  &  21.78 & 0.315 & $\cal{A}$ &
O47346\_4643  &  21.52   & star & $\cal{S}$ \nl
O47354\_4645   &  22.11 & 0.399 & $\cal{E}$,$\cal{I}$ &
O47376\_4706  &  23.03 & 0.400 &  $\cal{E}$ \nl
O47380\_4821   & 22.05  & 0.721 & $\cal{E}$ &
O47390\_4552   & 21.42  & 0.183 & $\cal{E}$,$\cal{I}$ \nl
O47390\_4603 &  21.42 & 0.578 & $\cal{A}$ &
O47393\_4351  &  21.25  & 0.348 &  $\cal{EI}$ \nl
O47410\_4253  &  20.87  & 0.348 & $\cal{A}$ &
O47411\_4340  &  19.70  & 0.348 & $\cal{A}$ \nl
O47419\_4449  & 21.12 & 0.539 & $\cal{E}$,$\cal{I}$ &
O47455\_4853  & 22.18    & 0.543 & $\cal{E}$,$\cal{I}$ \nl
O47480\_4514  & 21.55 & 0.361 & $\cal{E}$,$\cal{I}$ \nl
\enddata 
\tablenotetext{a}{Object names are based on their coordinates, 
Cxxyyy\_wwzz has the position 13 xx yy.y $-$11 ww zz, epoch J2000.} 
\tablenotetext{b}{The system of galaxy spectral types used is described
in Cohen \etal\ (1999).}
\tablenotetext{c}{This object has the spectrum of a M subdwarf.}
\label{table_cluster_nomem} 
\end{deluxetable}

%
%
\begin{deluxetable}{rrrr} 
\tablenum{3} 
\tablewidth{0pt} 
\tablecaption{Wavelengths for the Line Indices} 
\tablehead{
\colhead{Feature Name} & \colhead{Feature} & \colhead{Blue Continuum}
& \colhead{Red Continuum} \nl
\colhead{} & \colhead{(\AA)} & \colhead{(\AA)} & \colhead{(\AA)} \nl
\nl 
} 
\startdata 
3727\,\AA\ [OII] Emission & 3712 -- 3742 & 0.5(3713 -- 3741) & 
  0.5(3742 -- 3801) \nl
3933\,\AA\ Ca II Absorption & 3918 -- 3948 & 0.4(3500 -- 3670) & 
0.6(4030 -- 4090) \nl
D4000 (Break)\tablenotemark{a} &  ... & 3850 -- 3950 & 4000 -- 4100  \nl
\enddata
\tablenotetext{a}{This index is the ratio of the average flux in the
shorter wavelength bandpass to that in the longer.}
\label{table_bands} 
\end{deluxetable}

%
%
\begin{deluxetable}{llcccc} 
\tablenum{4} 
\tablewidth{0pt} 
\tablecaption{Mass Indicators For the Cluster 
of Galaxies \rxj} 
\tablehead{
\colhead{Reference} & \colhead{Method}
& \colhead{$\sigma$} & \colhead{$T_X$} & \colhead{Mass} &
\colhead{Radius} \nl
 &  &  \colhead{(\kms)} & \colhead{(keV)} &  
\colhead{(10$^{14}$ M$_\odot$)} & \colhead{ ($h^{-1}$ kpc)} \nl
} 
\startdata 
 & X-ray \nl
Schindler \etal\ (1997) & (ROSAT/ASCA) & [1320$\pm100$]& 9.3$\pm 
     1.0$& 2.9 / 8.5 & 500 / 1500 \nl
Allen (1998) & (ROSAT/ASCA) & [2500$^{+420}_{-800}$] & 26.4$^{+7.8}_{-12.3}$
 & 36$^{+11}_{-17}$ & 1000 \nl
Allen (2000) & (ROSAT/ASCA) & 1850$^{+270}_{-500}$ & 10.4 -- 26.4 & ... &  
    880 \nl
Ettori \etal\ (2001) & (BeppoSax) & [1635 -- 2250] & 13.2 -- 22.3 & ... & 
 1300\tablenotemark{a} \nl
Allen \etal\ (2002) & (Chandra) & [1545$\pm 50$] & 12.0$\pm 0.6$ & ... & 
    1000 \nl
 & Lensing \nl
Fischer \& Tyson (1997) &  weak lensing &  1500$\pm220$ & [11.5$\pm 2.8$]&  11$\pm3$ & 1000 \nl
Sahu \etal\ (1998) & strong lensing & 1300 & [9.1]&  3.4 & 140 \nl
This Paper & strong lensing &  850 / 1000 & [4.5 / 5.9] &  1.4 / 1.9 &  140 \nl
 & Galaxy $\sigma(v)$ \nl
This Paper & galaxy $\sigma(v)$ &  910$\pm130$ & [5.1$\pm1.2$] & 4.4$^{+1.4}_{-1.2}$ & $R_h=$380 \nl
\enddata
\tablenotetext{a}{This is the radius of the aperture used, but the
instrumental PSF is very broad.}
\label{table_sigmas} 
\end{deluxetable}

\clearpage

%

%

\end{document}